\documentclass{article}

\usepackage{PRIMEarxiv}

\usepackage[utf8]{inputenc} 
\usepackage[T1]{fontenc}    
\usepackage{hyperref}       
\usepackage{url}            
\usepackage{booktabs}       
\usepackage{amsfonts}       
\usepackage{nicefrac}       
\usepackage{microtype}      
\usepackage{lipsum}
\usepackage{fancyhdr}       
\usepackage{graphicx}       
\graphicspath{{media/}}     

\title{5G Network Security Practices: An Overview and Survey
\thanks{\textit{\underline{Citation}}: 
\textbf{Bannat Wala. M Kiran. 5G Network Security Practices: An Overview and Survey.}} 
}

\author{
  Fatema Bannat Wala \\
  ESnet \\
  Lawrence Berkeley National Laboratory \\
  Berkeley, CA\\
  \texttt{fatemabw@es.net} \\
   \And
  Mariam Kiran \\
  Quantum Communications and Networking \\
  Oak Ridge National Laboratory \\
  Oak Ridge, TN\\
  \texttt{kiranm@ornl.gov} \\
 }
 
\begin{document}
\maketitle

\begin{abstract}
This document provides an overview of 5G network security, describing various components of the 5G core network architecture and what kind of security services are offered by these 5G components. It also explores the potential security risks and vulnerabilities presented by the security architecture in 5G and recommends some of the best practices for the 5G network admins to consider while deploying a secure 5G network, based on the surveyed documents from the European government's efforts in commercializing the IoT devices and securing supply chain over 5G networks.
\end{abstract}



\section{5G Security Overview}
\begin{figure}
  \centering
    \includegraphics[width=0.85\linewidth]{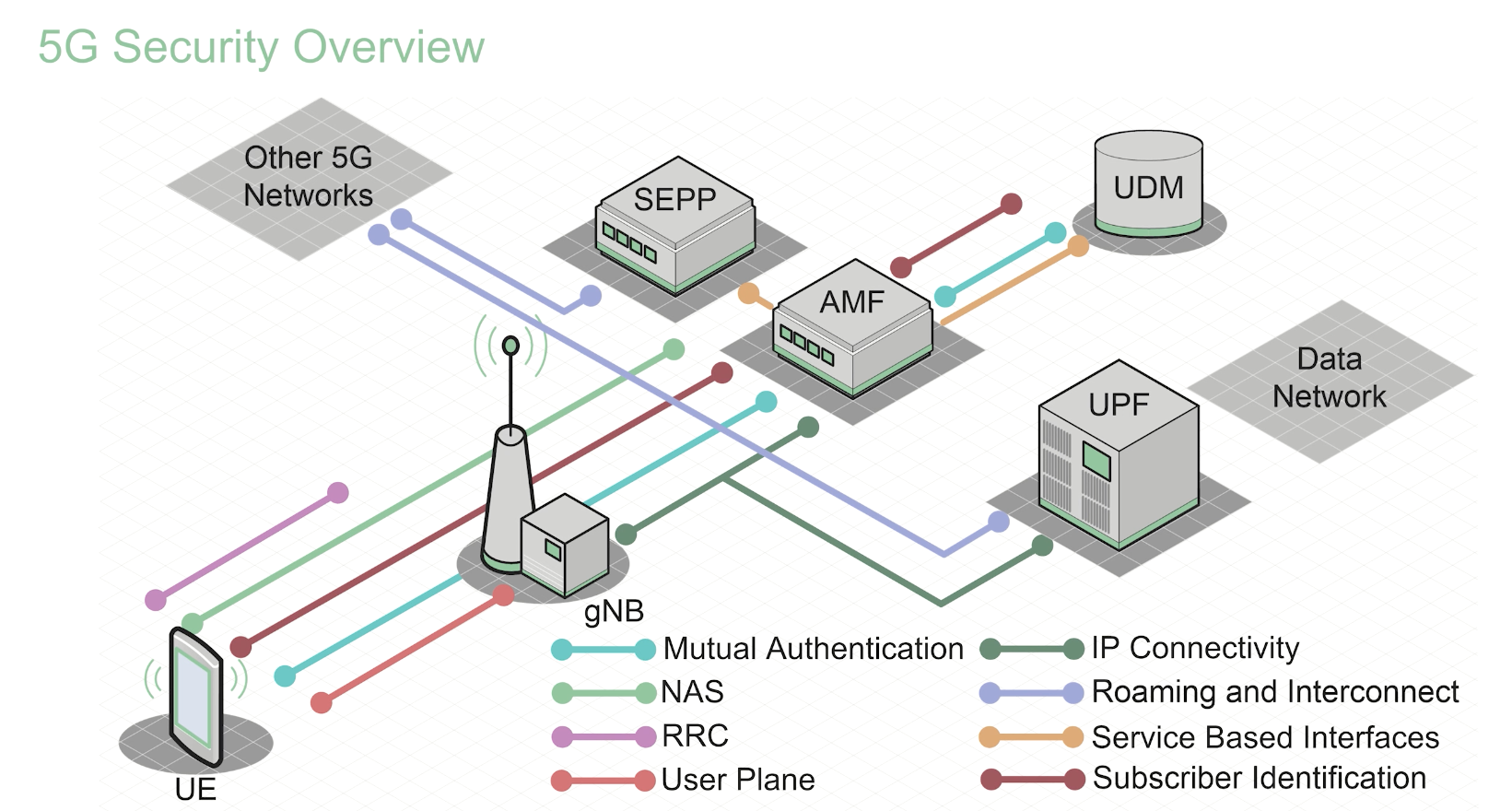}
    \caption{5G Security Overview}
  \label{fig:5gsec}
\end{figure}

The diagram in Figure \ref{fig:5gsec} shows various network functions that play a part in the overall 5G network security. The figure shows various elements that are somehow involved in providing the security in a 5G network. Starting with mutual authentication, it has been around in previous technologies before 5G and is also available in 5G, which happens between the user equipment (UE) and the Access and Mobility Management Function (AMF), but unified Data Management (UDM) also plays crucial role in the overall authentication process, as it hold the data associated with the UE identification. The detailed explanation of how it works is provided in the next sub-sections. Next, various types of signalling - RRC, NAS (non-access stratum) and User Plane traffic can be protected by the encryption and integrity protection options available in 5G. Encryption provides the confidentiality of the signalling messages transferred between UE/gNB and AMF, and integrity is provided by verifying the sender's and receiver's identity and making sure that the message is not forged in the transit. Although in all the signalling use-cases, confidentiality protection via encryption is an optional feature and not enabled by default. However, integrity protection in RRC and NAS signalling is mandatory and can't be disabled. Apart from the interaction of UE and AMFs, there are other components and network functions that come into picture when the UE tries to keep connectivity between other 5G networks while roaming, hence the roaming and inter-connections protection is provided by the security edge protection proxy (SEPP) and user plane functions (UPFs). The details on how SEPP provides protection for roaming connections is described in details in next sub-sections. AS we move into the network, we have got IP connectivity between RAN and the core, and sometimes that connectivity is not owned by the mobile service provider but some third party service provider, and in that case technologies like IPSEC can be used to protect that traffic.Lastly, the service based interfaces can also be protected using OAuth 2.0 and the necessary transport layer security, and can control which network functions are able to access services from other network functions. Finally, for the subscriber identification protection, 5G usually uses a temporary ID called 5G Global Unique Temporary Identifier (GUTI), but if there is a need for the UE to share it's IMSI over the radio network, that can also be protected using asymmetric encryption which is also called as SUCI (Subscriber Concealed ID). Next sub-sections describe in detail these fundamental methods of protecting 5G environment.

\subsection{Network Functions Virtualization NFVs and Network-Slicing}

The 3rd Generation Partnership Projects (3GPP) have defined 5G network to be a Service-Based Architecture (SBA), consisting of the various network functions (NFs) interconnected to deliver the control plane functionality and common data repositories of a 5G network. Since 5G is considered to be a SBA, a lot of it's NFs will be realized to be running in a virtualised environment to support the 5G core, the Radio Access Network (RAN) and Multi-access Edge Computing (MEC). Concepts like Cloud RAN (CRAN), Virtualised RAN (VRAN) and Open RAN (ORAN) to run baseband functions on commodity server hardware, are based on the Network Function Virtualization (NFV) principles.  This SBA model further adopts principles like modularity, reusability and self- containment of network functions to enable deployments to take advantage of the latest virtualisation and software technologies. NFV Management and Orchestration (MANO) covers the orchestration and lifecycle management of physical and/or software resources that support the virtualisation of the infrastructure, and the lifecycle management of VNFs.

Apart from NFVs being the base for the 5G service based architecture, network slicing is also another key feature introduced by 3GPP in 5G architecture. A network slice, within 5G scope, is considered as a set of 3GPP defined features and functionalities that together form a complete Public Land Mobile Network (PLMN) for providing services to UEs. Network slicing will allow customised control plane and user plane NFs to cater to the needs of a subset of users/UEs or related to the business customer needs. Because of which it is possible to create multiple network slices, each one composed of a collection of features, capabilities and services required to the needs of the slice. For example, one network slice can include the NFs to support mobile broadband services with full mobility support, and another one to support non-mobile, latency-critical industry applications.

\subsection{Mutual Authentication}
\begin{figure}
  \centering
    \includegraphics{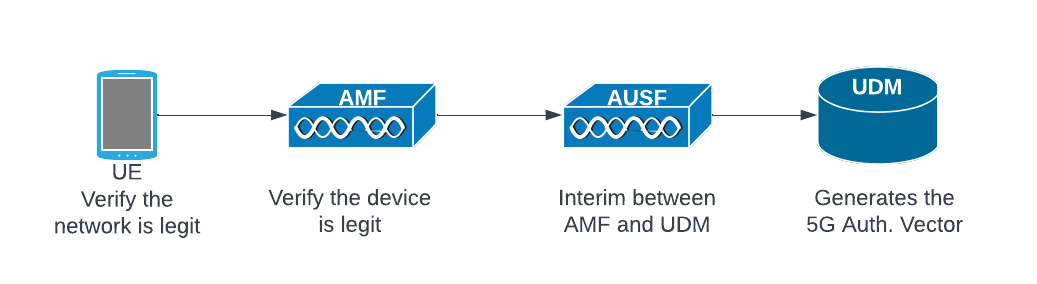}
    \caption{Mutual Authentication}
  \label{fig:MA}
\end{figure}

Mutual Authentication is not a new technique for 5G, it was actually introduced in 3G when it was recognized that the 2G algorithms had weaknesses around authentication. In 5G, however, before device tries to answer any security challenges coming from the network, it firstly verifies whether the network it is trying to connect to is legitimate. Once the device verifies the network, the device responds to the security challenge and allows the core network to verify that the device is legitimate. Usually this initial authentication verification happens between the UE and AMF, but the Authentication Server Function (AUSF) plays an important part in the mutual authentication and key agreement process, and it acts a liaison between the AMF and UDM to get the security information from UDM. The UDM generates the 5G authentication vector using the secret key of the device stored at the UDM, the same secret key is stored at the UE, hence the keys are symmetrical. This process of mutual authentication and the authentication vector creation are the first steps before any other security procedures are conducted in 5G. This mutual authentication process is shown in Figure \ref{fig:MA}.

\subsection{Encryption and Integrity}
\begin{figure}
  \centering
    \includegraphics{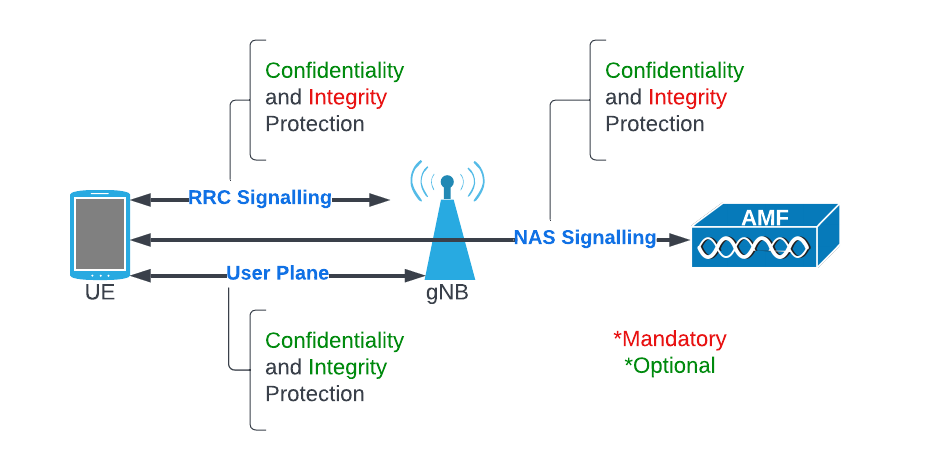}
    \caption{Encryption and Integrity}
  \label{fig:EI}
\end{figure}

Encryption and Integrity checking has been around since 2G, and 5G also provides the encryption and integrity checking for various types of signalling messages namely RRC, NAS and user-plane. These signalling require separate keys for the encryption and decryption processes and those keys are exchanged between the UE and AMF during the 5G mutual authentication and key agreement process which are distributed through the system where necessary. However not all these message signalling have same expectations in regards of encryption of messages. From the figure \ref{fig:EI}, it can be derived that for RCC and NAS signalling messages, encryption is optional and only integrity checking is mandatory and for the user-plane traffic both the encryption and integrity checking are optional, assuming that it depends on what type of device is connecting to the radio network and whether it would require encryption or not. If correctly used, these options of encryption of signalling messages for confidentiality protection between the user device and the network functions will potentially protect against the man in the middle (MITM) and fake base station attacks(Stingray/IMSI catcher).

\subsection{Service based interface protection} 
\begin{figure}
  \centering
    \includegraphics{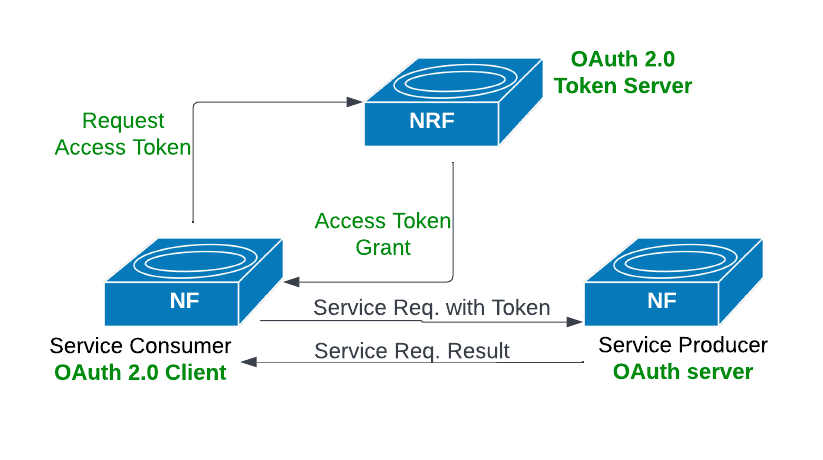}
    \caption{OAuth Token Based Authentication}
  \label{fig:OAuth}
\end{figure}

To protect the service based interfaces 5G provides two different options to consider. The first one is to protect the protocol stack for the service based interface by using transport layer security protocol (TLS) between HTTP/2 and TCP. Only TLS v 1.2 and above are allowed to be used, furthermore 3GPP also specifies different cipher suits that are allowed to be used in the TLS security. TLS provides integrity and encryption of all the service based interface messages in transit. The other method available is to use OAuth 2.0, which doesn't necessarily provide protection of traffic in transit but provides the protection of service producer against malicious service consumers trying to inject malicious attempts to invoke a service at the service producer. OAuth 2.0 is token based authentication method, where the network repository function acts as OAth 2.0 token server and a network function acting as an OAuth 2.0 client requests for the access token to be granted to access a specific service at a network function acting as OAuth 2.0 server. The access tokens can be applicable to a specific type of network function or can be much more granular than that, and can also be applied to a very specific network function. The token can grant very granular permissions to access to a single service or service operations or a group of services, which can then be part of the service request sent by the client (part of the HTTP/2 header) to the server. The server verifies the authenticity of the token and checks if it is valid and then grants the access to the requested service. These both methods of providing TLS security and token based authentication are optional. Figure \ref{fig:OAuth} shows the basic OAuth between the service consumer and the service provider. 

\subsection{Roaming and inter-connections security}
\begin{figure}
  \centering
    \includegraphics{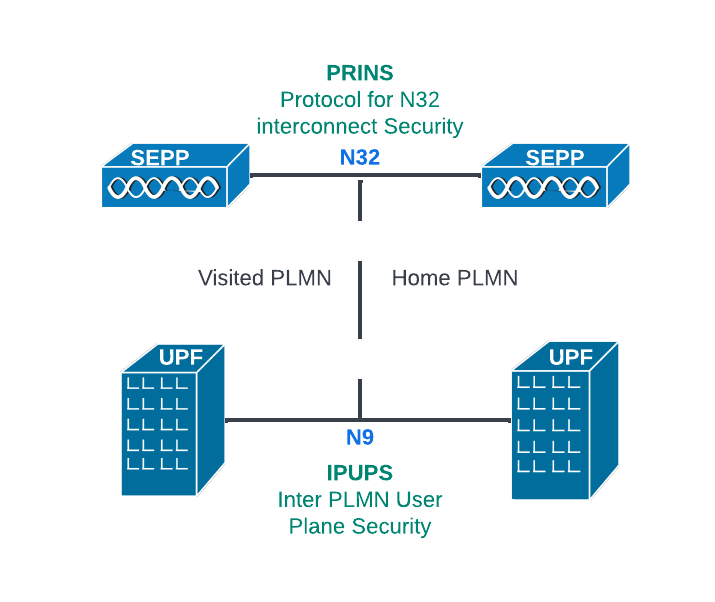}
    \caption{Roaming and inter-connections security}
  \label{fig:RI}
\end{figure}

In most cases we don't have direct connectivity between the visited Public land mobile network (PLMN) and home PLMN. Usually the network connection transition between Visited PLMN and Home PLMN is handled by IPX (IP packet exchange), and for the obvious reasons it is not owned by the mobile network service provider, rather a third party connection provider. Hence, it's important to consider the security of the traffic as it traverses between different network interfaces. In service based architecture that security is provided by Security Edge Protection Proxy (SEPP) for the control plane traffic. It has number of security features like topology hiding support, where any sensitive addressing information is filtered out in any signalling messages, message filtering supporting filtering of any specific messages and traffic policing where traffic streams can be policed. As shown in the figure \ref{fig:RI}, these SEPPs are connected by N32 protocol, which is an enhancement over older 4G S6a protocol, in a way that it provides the TLS protection under HTTP/2, parts of which can also be protected using PRINS (Protocol for N32 Interconnect Security) which allows one to selectively protect parts of the message, leaving other parts that are required for end to end routing in the inter-connected network. For user plane function (UPF), Inter PLMN User Plane Security (IPUPS) is used which provides N9 protocol connectivity between UPFs. There will be no traffic passing across N9 unless it is associated with a pre-configured GTP v-1u tunnel which prevents any unintentional/malicious traffic from traversing one UPF to reach other UPF if a GTP tunnel hasn't been setup and is unavailable, hence that traffic is dropped. Both the UPFs (each end of the connection) have to be aware of the traffic between each other.

\subsection{Subscriber Identity protection}

\begin{figure}
  \centering
  \includegraphics{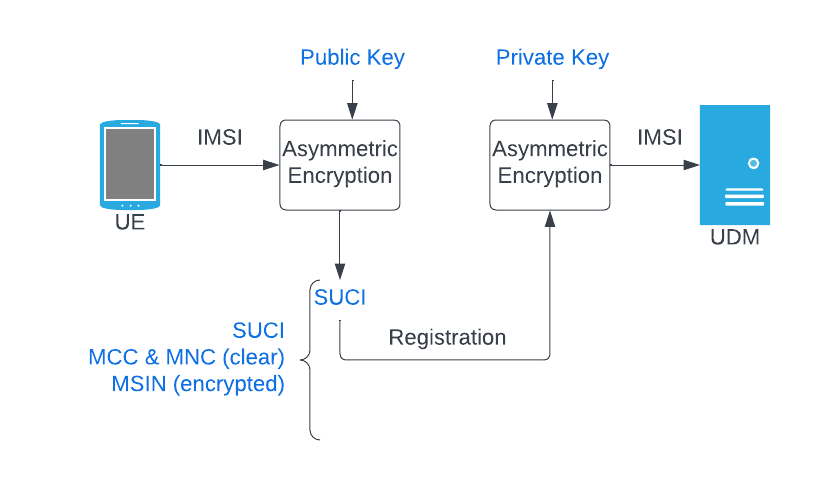}
  \caption{Subscriber Identity Protection}
  \label{fig:SUCI}
\end{figure}

In 5G a mechanism called as Subscription Concealed ID (SUCI) is introduced, which means that the device (UE) never has to send IMSI(in 5G terminology its called subscription permanent ID) in clear text. The user element converts IMSI to SUCI by encrypting the MSIN (Mobile Subscriber Identification Number) using the public key derived from the 5G network itself, leaving MCC (Mobile Country Code) and MNC (Mobile Network Code) in clear as it's required for the routing purposes. During the registration process the UE sends over the SUCI to the UDM where it gets decrypted by the UDM using the corresponding private key and the IMSI is derived. It is not mandatory however to use encryption of MSIN in SUCI and the algorithm can be set to null for clear text SUCI. The figure \ref{fig:SUCI} shows how SUCI is transferred between UE and UDM.

\section{5G security best practices and recommendations}

Most of these recommendations are inspired by the European Union Agency for Cybersecurity 
(ENISA). The ENISA is the Union's agency dedicated to achieving a high common level of cybersecurity across Europe.

The best practices can mainly be categorized into three important areas, policy, technical and organizational. The details of these practices are described in the Challenges and Best Practices document by ENISA. Below sub-sections highlights only critical and most important ones. For the complete list, please refer to the ENISA doc \cite{72}.

\subsection{Policy based}
Zero-Trust:
5G NFV deployments should build on mature cybersecurity standards employed in enterprise and cloud environments. Examples include the NIST Cybersecurity Framework and ISO Information Security Management System series71, the ETSI GS NFV-SEC 00372 Security and Trust Guidance and the UK NCSC Zero Trust Architecture Design Principles \cite{73}.
Zero Trust represents an overarching access security model that deliberately avoids assuming implicit trust between elements in a network. This is particularly important in 5G as various external stakeholders may need to access infrastructure components or services for management, maintenance, or monitoring purposes \cite{74}.

Vulnerability handling and patch management:
NFV and MANO software components will need to be monitored for vulnerabilities and patched as quickly as possible to address evolving risks and ensure security and functionality.

Security assessment of new or changes to existing VNF service templates:
New or modified VNF service templates should be validated through proper risk assessment by a security professional.

Security testing and assurance:
Regular penetration and vulnerability testing should be performed across the NFVi and MANO production environment to identify any vulnerabilities (e.g. OSs, hypervisor, CIS, VMs or containers) or compromise of the network zoning rules. 5G stakeholders should also leverage internationally recognised product testing, assurance, and certification regimes. Potential solutions to these requirements include, among others, the Network Equipment Security Assurance Scheme (NESAS) \cite{76}, jointly defined by 3GPP and GSMA, and the ENISA EUCC\cite{77}, EUCS\cite{78} schemes.

\subsection{Organizational based}
Secure physical environment and geographical location:
Security is not just about encrypting data streams but also about the physical deployment of equipment. NFV devices and equipment location, type of equipment, and type of services running on the equipment are all parts of the complete system and must be protected in a secure physical environment.

Trust model:
A trust model of relationships among various 5G stakeholders should be built and be able to answer questions such as: `for what one does on trust?', `How much should one trust?' and `How much anyone can trust?'.

SLAs establishment:
SLAs established between 5G stakeholders should include important security and compliance measures. In this respect, SLAs are like an insurance policy for the 5G network services.
SLAs set the expectations for the performance of a service provider and for the security level. They also establish penalties for missing targets. In that concept, NFV could not leave unaffected the evolution of SLA models and their flexibility in adapting to more demanding parameters. An SLA management framework should be defined in order to fill the gap between 5G stakeholders.

\subsection{Technical based}
Security monitoring and filtering:
This covers the use of virtual network security appliances such as anti-virus, virtual firewalls or virtual IDS/IPS to achieve a level of security comparable to traditional networks. Further, machine learning (ML) assisted solutions can be used to detect attack traffic (e.g. DoS attacks) and distinguish it from normal traffic, so that it can be handled appropriately.

Data protection and privacy:
To prevent disclosure of user data (e.g. subscriber identifiers), 5G networks should use security mechanisms such as the encryption cryptographic operation. It is important to pay attention to the requirements of lawful interception60. Privacy and security, both of individuals' personal data and of critical infrastructure, are important preconditions for GDPR. Encryption is a crucial tool to achieve these goals. Any approach to weaken or grant backdoor access to encryption methods defeats the entire purpose of encryption and undermines users' trust, exposing 5G systems to increased risks. At the same time, it remains vitally important that companies and law enforcement authorities continue to work together, ensuring that authorities have the best methods and access to electronic evidence without weakening or putting strong encryption at risk.

Centralised log auditing:
All the NFV, SDN and MANO elements should submit information on security events to a centralised platform, which shall monitor and analyse the logs in real time for possible attempts at intrusion.

\section{Security risks and vulnerabilities}

There are two major 5G deployment models so far, although there will be more options added in future. The currently deployed model is called non-standalone (NSA) mode, more precisely referred to as EN-DC. In this model, the 5G stations are integrated with an existing 4G network working in tandem with LTE base stations and hence connected to the LTE core, which relies on the measures and security protections of 4G/LTE network. Again, this cause a major security risk implication in 5G because of the reliance on the 4G core network. This backward compatibility with current deployment models of 5G will pose a security risk in terms of the security vulnerabilities inherited from 4G network. The next phase of 5G deployment will most likely be Stand Alone (SA) mode, more precisely referred to as SA-NR. As the name suggest, in this model, 5G new radio network (NR) will be connected directly to the 5G core, which will allow the full security features of 5G specifications to be realised. Although it is recognised that new paradigms (cloud native, service-based architecture) will introduce new security challenges.

Apart from the enhanced security features supported by 5G, there will be new attack vectors and threat models which will be realized as 5G's adaptation will expand to the AI devices (intelligent cars etc.), Industrial control devices, and most importantly the IoT sector, which will include every corner case of a low level device capable of connecting to the internet. Internationally, 5G is already being deployed commercially to the supply-chain industry. Because of this broad paradigm that 5G open up to the internet, and given the fact that historically the telecommunication and mobile network service provides paid less attention to the security while building custom protocol and software for the mobile/wireless communication stack, it will be daunting task for 5G security to keep up with the industry grade standard of providing security and privacy services. Also, not to mention that wireless being it's own bubble, developing custom software stacks and protocol, the attackers might not be very well versed to exploit it efficiently. Hence, there is a learning curve there as well, which balances out the 5G security risks, but it won't be long until adversaries would be able to catch up with the wireless technology and will start exploiting potential vulnerabilities. Therefore, precautions should be taken from the start to build a better and more secure 5G network and close as many gaps as possible.

\section{Conclusion}
5G is an opportunity for the mobile industry to enhance network and service security. New authentication capabilities, enhanced subscriber identity protection and additional security mechanisms will result in significant security improvements over legacy generations.

Experience has shown that 2G/3G networks make use of insecure, un-managed protocols and are subject to fraud and threats on a regular basis. Many of these attacks have been mitigated with 4G and 5G. However, due to the backward compatibility of 4G with 3G/2G they will not disappear until the legacy technology or backward compatibility is ceased.

When defining 5G roll-outs operators will have to consider how these legacy networks will impact them over time; considering how attacks could be prevented if legacy generations are either isolated or removed from the ecosystem.

With the advent of new transport protocol for fast and secure communication, QUIC - Quick UDP Internet Connections, and 5G's stepping into the network protocol stack more, it would be worth while to investigate the performance impact of implementing 5G over QUIC, which might be used as default if 5G will run HTTP3 for service based interfaces. This area is currently being explored and the performance and security enhancement will be tested for this experiment.

Lastly, since the industry is moving towards Zero-Trust implementation, we are also currently exploring and researching how 5G deployment models will fit in the Zero Trust Architecture, as 5G once deployed will constitute a part of the network where users will be connecting to network to access the internet and and authentication and authorization will come into play together with other major pillars of ZTA.

\section*{Acknowledgments}
This work is funded under DOE Advanced Scientific and Computing Research Program contract number ERKJ437 for Self-Driving 5G for Science.
\bibliographystyle{unsrt}  
\bibliography{references}

\end{document}